\documentclass[RNAAS]{aastex631}
\usepackage{natbib}
\usepackage{amsmath,upgreek}
\usepackage{graphicx}%
\usepackage{multirow}%
\usepackage{amsmath,amssymb,amsfonts}%

\begin{document}

\title{How stable are the impact factors of major astrophysics journals?}

\author[0009-0000-0039-6736]{Rayani Venkat Sai Rithvik} 
\affiliation{Department of Electrical Engineering,
IIT Hyderabad,   Telangana-500084 India}
\author[0000-0002-0466-3288]{Shantanu Desai}
\affiliation{Department of Physics, IIT Hyderabad,   Telangana-500084 India}

\correspondingauthor{Shantanu Desai}
\email{shntn05@gmail.com}

\begin{abstract}
We examine the long-term trends in impact factor over the past decade (2015–2024) for seven flagship journals widely used in astrophysics: ApJ, AJ, ApJL, ApJS, A\&A, JCAP, and MNRAS.
 We checked the variation of both the traditional impact factor and the median-based impact factor, which we had studied in a previous work. We find that ApJS exhibits the largest drop in impact factor from 14.9 to 9.0, while ApJ and MNRAS show a steady decline from 6.7 to 5.2 and 5.6 to 5.0, respectively. However, the impact factors for AJ and ApJL have shown a steady increase from 4.55 to 5.35  and 6.4 to 9.7, respectively. The median-based impact factors are much more stable and show variations of at most $\pm 1$ within the last decade.

 \end{abstract}





\section{Introduction}
In a previous work~\citep{Rithvik24} (R25 hereafter), we have estimated  the impact factors of about 40 Astrophysics journals  using NASA/ADS~\citep{ADS}. We also defined a new median-based impact factor and compared the two metrics for each of the journals. We briefly recap the definitions of these impact factors.

The  traditional way to calculate the impact factor, (which we referred to as Old Impact Factor in R25) in year \( n \) is defined as the ratio of the total number of citations in year \( n-1 \) of all papers published in the journal during years \( n-2 \) and \( n-3 \), divided by the number of refereed papers published in those same years~\footnote{\url{https://en.wikipedia.org/wiki/Impact_factor}}:

\begin{equation}
\text{Impact Factor}_{\text{old}}(n) = \frac{C_{n-1}}{P_{n-2} + P_{n-3}}
\label{eq:IF}
\end{equation}

where:
\begin{itemize}
    \item \( C_{n-1} \) is the total number of citations in year \( n-1 \) to papers published in years \( n-2 \) and \( n-3 \).
    \item \( P_{n-2} \) is the number of refereed publications in the journal in year \( n-2 \).
    \item \( P_{n-3} \) is the number of refereed publications in the journal in year \( n-3 \).
\end{itemize}
Note that the numerator in Eq~\ref{eq:IF} does not include the citations published in the same year as the journal publication. The citations consist of  both refereed and unrefereed publications.  We note that the official impact factors are calculated by Clarivate~\footnote{https://mjl.clarivate.com/home},   and are based on the citations obtained using Web of Science bibliometric data.


In R25, we introduced  a new impact factor based on the  median number of citations, which we refer to as  \textbf{New Impact Factor}.
The \textbf{New Impact Factor} in year \( n \) is defined as the median number of citations in year \( n-1 \) to all the refereed papers published in the journal during years \( n-2 \) and \( n-3 \), divided by the total number published papers during years \( n-2 \) and \( n-3 \). 

In this work, we extend the analysis in R25, by studying the trends of both the old and new impact factors over the last decade  for  seven of the most widely used research journals in Astrophysics: ApJ (Astrophysical Journal), AJ (Astronomical Journal), ApJL (Astrophysical Journal Letters), ApJS (Astrophysical Journal Supplement), MNRAS (Monthly Notices of Royal Astronomical Society), A\&A (Astronomy and Astrophysics), as well as JCAP (Journal of Cosmology and Astroparticle Physics). 
This would help us assess the stability of the impact factor.
The first six journals have long been considered as  the flagship journals in astrophysics, with some of them (ApJ, MNRAS) publishing papers as early as the 19th century. JCAP is an online peer reviewed journal that has been publishing papers since 2003 and is widely used by the Cosmology and Particle Astrophysics community, and has an impact factor comparable to the flagship journals used in Astrophysics.
The manuscript is structured as follows. Our results are described in Sect.~\ref{sec:results} and we conclude in Sect.~\ref{sec:conclusions}.

\section{Results}
\label{sec:results} 
Similar to R24, we collated  citations using NASA/ADS API available at \url{https://ui.adsabs.harvard.edu/help/api/} and checked for Astronomy publications by using the tag ``collections:Astronomy''. We have shown in R24 that the NASA/ADS-based impact factors do not always agree with the official impact factors.

\begin{figure}
    \centering
    \includegraphics[width=0.49\linewidth]{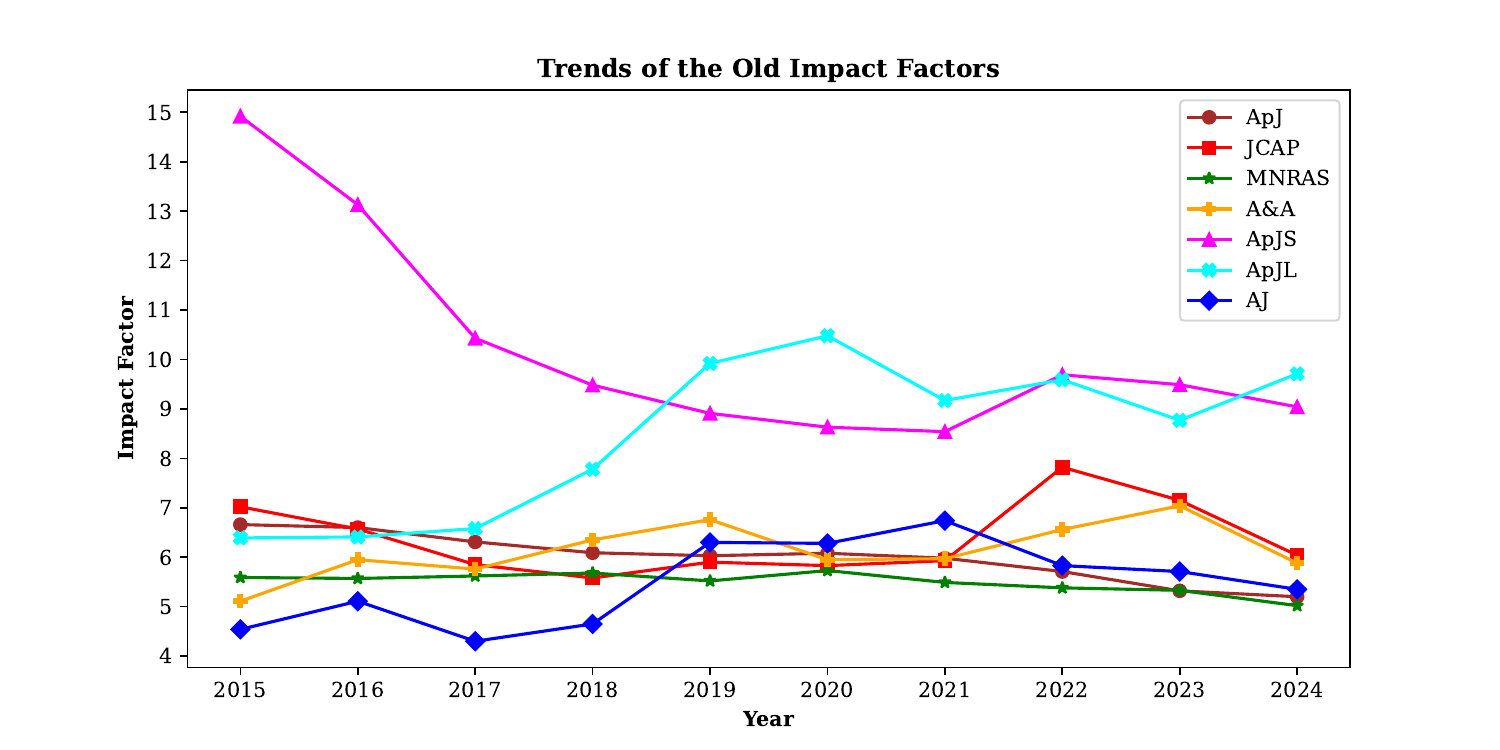}
    \includegraphics[width=0.49\linewidth]{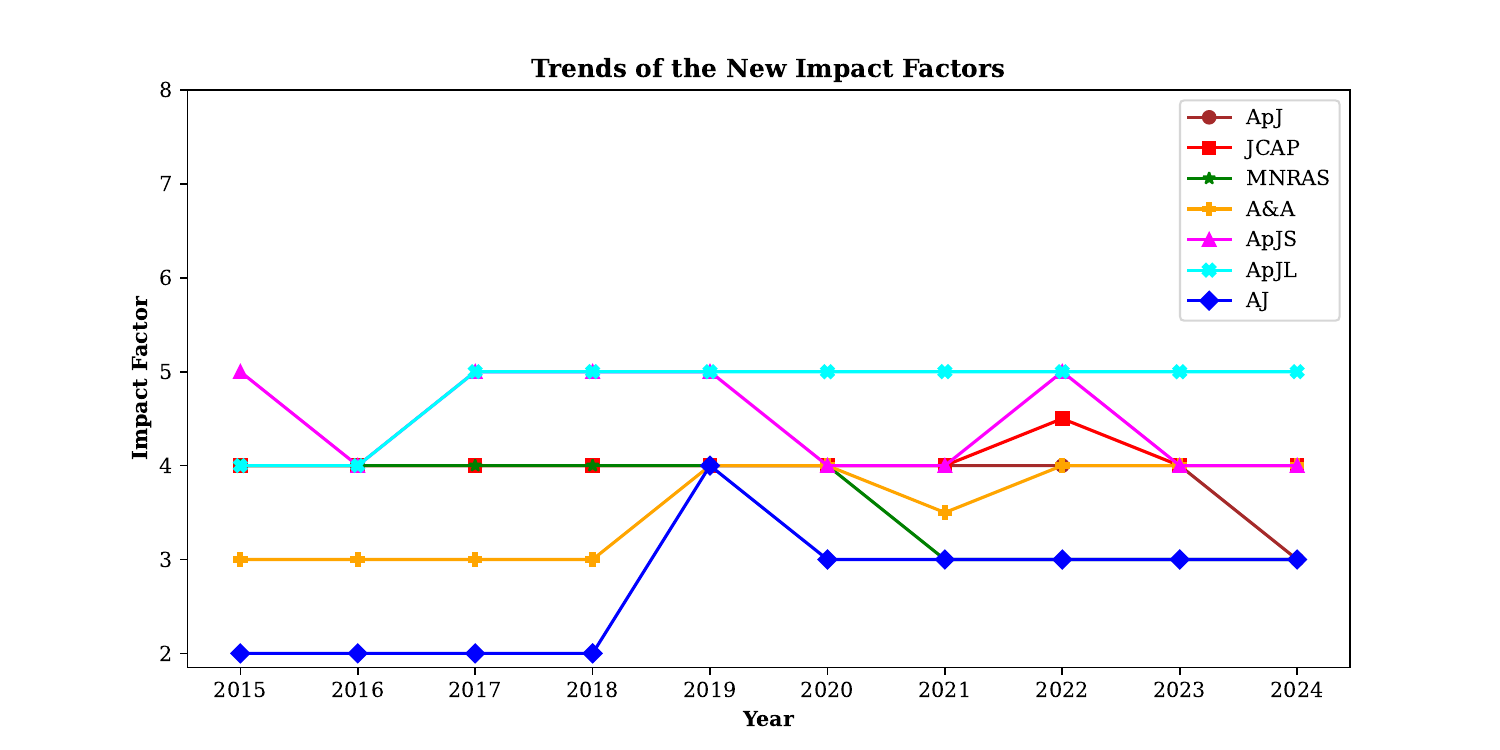}
    \caption{Trends of the old  and   new (median-based)  impact factor of seven astrophysics journals from 2015-2024 in the left and right panels respectively.}
    \label{fig:IF}
\end{figure}

Our results can be found in tabular form in Table~\ref{tab:tab1}. The trends of old and new impact factor can be found in Fig.~\ref{fig:IF} in the left and right panel, respectively.
Some of the key features based on the results from the old  impact factor  are as follows:
\begin{itemize}
\item The greatest decrease in impact factor  is observed for ApJS, which has fallen from $\sim$ 14.9 (in 2015) to 9.0 (in 2024). 
\item ApJ has shown a steady decline in the impact factor from 6.66 (in 2015) to 5.2 in (2024). 
\item A similar steady decline was observed for MNRAS from 5.62 to 5.02.
\item The impact factors for ApJL and AJ have steadily increased, rising    from 4.55 to 5.35 for AJ, and 6.40 to 9.71 for ApJL.
\item The impact factor for A\&A has fluctuated the most with two peaks in 2019 and 2023 followed by a decline. Overall, there has been an increase in the IF of  about  0.77  from 5.11 to 5.88. The increase in impact factor between 2018 and 2023 is likely due to papers from the Planck collaboration. 
\item Similarly, JCAP impact factor showed  a steady decline from 2015 (7.02)  to 2020 (5.83), followed by a steady rise until 2022 (7.82), followed by another drop last two years with the 2024 impact factor equal to 6.04.
\end{itemize}
Compared to the old impact factor, the new impact factors for these journals have been much more stable. We now highlight some of the salient features.
\begin{itemize}
\item The impact factor for ApJL increased by one from 4 to 5 in 2017 and has remained constant since then.
\item The impact factor for AJ shows  a maximum (4)  in 2017 (increasing  by 2 compared to 2015), and then decreasing by 2 from 2018 onwards and remaining constant since then.  
\item The impact factor for ApJS has oscillated between 4 and 5 in the last 10 years.
\item The impact factor for A\&A increased from 3 to 4 in 2019 and has been constant since then, except for one dip in 2021, when it dropped to 3.5
\item The impact factor for MNRAS was constant (4) until 2021, when it decreased by one (3), and has remained constant since then.
\item The impact factor for JCAP has remained constant with a value of 4 throughout the last decade, except for 2022, when it increased to 5.
\item The impact factor for ApJ has remained constant (4) until 2023 and then decreased to 3 in 2024. 

\end{itemize}

\begin{table*}[t]
    \centering
    \begin{tabular}{|c|c|c|c|c|c|c|c|c|c|c|}
        \hline
        & \textbf{2024} & \textbf{2023} & \textbf{2022} & \textbf{2021} & \textbf{2020} & \textbf{2019} & \textbf{2018} & \textbf{2017} & \textbf{2016} & \textbf{2015} \\
        \hline
        \textbf{ApJ}  & 5.20/3  & 5.32/4  & 5.71/4  & 5.98/4  & 6.08/4  & 6.03/4  & 6.09/4  & 6.31/4  & 6.60/4  & 6.66/4  \\
        \hline
        \textbf{MNRAS} & 5.02/3  & 5.33/3  & 5.38/3  & 5.49/3  & 5.73/4  & 5.52/4   & 5.68/4  & 5.62/4  & 5.57/4  & 5.59/4  \\
        \hline
        \textbf{A\&A}  & 5.88/4  & 7.08/4  & 6.57/4  & 5.97/3.5  & 5.95/4  & 6.76/4  & 6.36/3  & 5.76/3  & 5.96/3  & 5.11/3  \\
        \hline
        \textbf{JCAP}  & 6.04/4  & 7.15/4  & 7.82/4.5  & 5.93/4  & 5.83/4  & 5.90/4  & 5.58/4  & 5.85/4  & 6.57/4  & 7.02/4  \\
        \hline
        \textbf{ApJS}  & 9.04/5  & 9.49/4  & 9.70/5  & 8.56/4  & 8.63/4  & 8.91/5  & 9.50/5  & 10.44/5  & 13.15/4  & 14.92/5  \\
        \hline
        \textbf{ApJL}  & 9.71/5  & 8.78/5  & 9.60/5  & 9.18/5  & 10.49/5  & 9.93/5  & 7.78/5  & 6.58/5  & 6.42/4  & 6.40/4  \\
        \hline
        \textbf{AJ}    & 5.35/3  & 5.71/3  & 5.83/3  & 6.74/3  & 6.28/3  & 6.31/4  & 4.66/2  & 4.31/2  & 5.11/2  & 4.55/2  \\
        \hline
    \end{tabular}
    \caption{Journal impact factors over the years, in the format of old impact factor/new impact factor evaluated using NASA/ADS. Note that old impact factor refers to the traditional way of calculating impact factor  (cf. Eq.~\ref{eq:IF}) and new impact factor refers to the median based impact factor.}
    \label{tab:tab1}
\end{table*}

\section{Conclusions}
\label{sec:conclusions}
In this work, we have looked at the stability of the traditional (old) and median-based (new) impact factors of seven flagship journals in Astrophysics and Cosmology over the last decade (from 2015-2024). For this purpose, similar to our previous work we use the citations collated by NASA/ADS. A tabular summary of our results can be found in Table~\ref{tab:tab1}, whereas a  graphical summary can be found in Fig.~\ref{fig:IF}.

We find a lot of variations in the old impact for most journals. The largest decrease has been observed for ApJS, dropping  from 14.9 to 9.0 over  the past decade. The old impact factors for ApJ and MNRAS  have also steadily declined from 6.7 to 5.2 and 5.6 to 5.0, respectively. The remaining journals have exhibited  fluctuations in their values with a net increase from 4.55 to 5.35, 6.4 to 9.7, for AJ and ApJL, respectively, along with a net decrease from  7.0 to 6.0 for JCAP. The results for the new impact factors have been much more stable with maximum variations of $\pm 1$ in the past decade.

\section*{Acknowledgments}
We are thankful to  Choong Ngeow, Noam Soker, Alex Saro, and Zhaozhou Li for encouraging comments on R25, which motivated us to carry out this study.


\bibliographystyle{aasjournal}
\bibliography{Wiley-ASNA}%
\end{document}